TITLE: Interfertile oaks in an island environment. II. Limited hybridization between *Quercus alnifolia* Poech and *Q. coccifera* L. in a mixed stand.

AUTHORS: Charalambos Neophytou*[1,2], Filippos A. Aravanopoulos[2], Siegfried Fink[3], Aikaterini Dounavi[1]

CORRESPONDING AUTHOR:
Charalambos Neophytou
Forest Research Institute of Baden-Württemberg
Department of Forest Ecology
Wonnhaldestr. 4
D-79100 Freiburg
Germany

E-mail address: Charalambos.Neophytou@forst.bwl.de
Current e-mail address: chneophytou@gmail.com
Telephone: +49 / 761 / 40 18 159
Fax: +49 / 761 / 40 18 133

The final publication is available at:
http://link.springer.com/article/10.1007/s10342-010-0454-4#page-1

[1] Department of Forest Ecology, Forest Research Institute of Baden-Württemberg, Wonnhaldestr. 4, D-79100, Freiburg, Germany
[2] Laboratory of Forest Genetics and Tree Breeding, Faculty of Forestry and Natural Environment, Aristotle University of Thessaloniki, P.O. Box 238, Thessaloniki, Greece
[3] Chair of Forest Botany, Faculty of Forest and Environmental Sciences, Albert-Ludwigs University of Freiburg, Bertoldstr. 17, 79085, Freiburg, Germany

ABSTRACT

Hybridization and introgression between *Quercus alnifolia* Poech and *Q. coccifera* L. is studied by analyzing morphological traits, nuclear and chloroplast DNA markers. The study site is a mixed stand on Troodos Mountains (Cyprus) and the analyzed material includes both adult trees and progenies of specific mother trees. Multivariate analysis of morphological traits shows that the two species can be well distinguished using simple leaf morphometric parameters. A lower genetic diversity in *Q. alnifolia* than in *Q. coccifera* and a high interspecific differentiation between the two species are supported by an analysis of nuclear and chloroplast microsatellites. The intermediacy of the four designated hybrids is verified by both leaf morphometric and genetic data. Analysis of progeny arrays provides evidence that interspecific crossings are rare. This finding is further supported by limited introgression of chloroplast genomes. Reproductive barriers (e.g. asynchronous phenology, post-zygotic incompatibilities) might account for this result. A directionality of interspecific gene flow is indicated by a genetic assignment analysis of effective pollen clouds with *Q. alnifolia* acting as pollen donor. Differences in flowering phenology and species distribution in the stand may have influenced the direction of gene flow and the genetic differentiation among effective pollen clouds of different mother trees within species.

INTRODUCTION

*Quercus alnifolia* and *Quercus coccifera* are the two representatives of sclerophyllous oaks in Cyprus. They are interfertile and there is no genetic evidence that they hybridize with the phylogenetically distant *Q. infectoria* ssp. *veneris*, the third oak species which is present on the island (Neophytou et al. 2008). Recent studies were stimulated by observations of individuals with intermediate morphology, well known among local people. Intermediates have been morphologically described (Hand 2006) and leaf morphometric traits of the two species and their potential hybrids have been surveyed at the population level (Neophytou et al. 2007). More recently, molecular differentiation between these species was studied at a large scale indicating that the two species constitute two well separated units in terms of nuclear DNA differentiation. In addition, analysis of several populations indicated that they widely share their chloroplast genomes (Neophytou et al. 2010a). Here, we seek to analyze hybridization patterns in a mixed stand intensively, by surveying both morphological and genetic differentiation.

Leaf morphology has been widely used to distinguish between related oak species. Several different approaches have been used, varying from simple observation of diagnostic traits to univariate and multivariate methods (see Rushton 1993 for a review). In general, multivariate analyses have been established as a powerful and easily applicable tool for studying differentiation and hybridization in the Fagaceae, e.g. in oaks (Kremer et al. 2002) and chestnut (Aravanopoulos 2005). For *Q. alnifolia* and *Q. coccifera*, multivariate analysis with use of simple leaf measures has been efficiently used to discriminate between the two parental species and to detect introgressed forms (Neophytou et al. 2007). However, conclusions based only on morphology are limited. This is more difficult when hybrids of the second or higher generations are involved. Backcrossed individuals often regain the morphological identity of the parental species (Rubio de Casas et al. 2007). Moreover, phenotypic plasticity can lead to a higher morphological variability. This can lead to an overestimation of the gene pool or to phenotypic overlapping that is not connected with genetic introgression (Jimenez et al. 2004).

The development of several molecular markers in the last two decades has opened new perspectives in the study area of hybridization and introgression in oaks. Research has largely focused on economically important species, like *Q. petraea* and *Q. robur* in Europe. Two large categories of DNA markers have been used to answer different questions. On one hand, nuclear DNA markers – mainly highly variable microsatellites – have been applied to study genetic differentiation, genetic structures and gene flow between related species (e.g. Muir et al. 2000; Gugerli et al. 2007; Lepais et al. 2009; Neophytou et al. 2010b). On the other hand, plastid DNA

markers – mostly from the chloroplast genome – have been used for studying phylogeny, postglacial recolonization patterns and historic genetic introgression (e.g. Petit et al. 2002; Finkeldey and Matyás 2003). Results have provided important insights into the issue of hybridization in oaks and at the same time revealed the complexity of this phenomenon. For example, it has been shown that rate and directionality of interspecific gene flow vary strongly depending among others on the phylogenetic affinity of the studied oak species (Rushton 1993), the presence of reproductive barriers (Boavida et al. 2001), the variability of ecological site conditions leading to adaptive barriers (Dodd and Afzal-Rafii 2004), the relative abundance of each hybridizing species (Lepais et al. 2009) etc.

A contrasting pattern between nuclear and chloroplastic differentiation has been shown between *Q. alnifolia* and *Q. coccifera* in Cyprus. On one hand, the high levels of nuclear differentiation (e.g. multilocus interspecific $F_{ST}$ between 0.310 and 0.364) support a limited degree of hybridization. On the other hand, shared chloroplast DNA (cpDNA) structures which are regionally distributed suggest that the two species have exchanged their chloroplast genomes through hybridization and backcrosses. The aforementioned results are presented in a companion paper, which investigates genetic differentiation between *Q. alnifolia* and *Q. coccifera* at a large scale by comparing pure populations of the two species, as well as one mixed stand (Neophytou et al. 2010a). In the present communication, we aim to elucidate the extent of contemporary hybridization between *Q. alnifolia* and *Q. coccifera* in sympatry by explicitly focusing on the mixed stand included in Neophytou et al. (2010a). For this purpose, we combine data from leaf morphology, nuclear and chloroplast DNA of all adult trees of the stand. Additionally, we analyze the effective pollen clouds of known maternal trees representing both species and intermediates, in order to investigate the degree and directionality of interspecific gene flow.

MATERIALS AND METHODS

*The study species*

*Quercus alnifolia* is an evergreen shrub or a much branched wide-crowned small tree up to 10 m, exceptionally reaching 14 m under optimal conditions (Meikle 1977, Knopf 2006). It is characterized by its dark green glabrous leaves with their golden tomentous lower surface. *Quercus alnifolia* grows exclusively on the igneous rock formations of the Troodos Mountains in Cyprus in altitudes between 400 and 1800 m. Flowering occurs between the end of April and the beginning of June, depending on the altitude. Acorns display an annual maturation cycle (Knopf 2006).

*Quercus coccifera* is an evergreen shrub or small tree up to 10 m, occasionally attaining large heights up to 20 m (Meikle 1977; Chatziphilippidis 2006). Its leaves are leathery with serrate margins and glabrous or thinly stellate pubescent lower surface. In Cyprus, it can be found in a wide variety of ecosystems occurring in altitudes from near sea level up to 1400 m. Flowering takes place between April and May. Acorn maturation is predominantly biennial, although more complex reproductive cycles involving also annual acorn maturation have been observed (Bianco and Schirone 1985).

*Sampling*

A mixed stand of *Q. alnifolia* and *Q. coccifera* in the western part of the Troodos Mountains was selected. The stand is located east of the village of Kambos, on a west exposed slope at an altitude of 700-800 m (coordinates: 35°02′N, 32°44′E). The geological substrate consists of diabasic rocks, belonging to the ophiolite geological formation of Troodos. The area had been partly used agriculturally until the early 80s and had been subjected to intensive grazing by goats until 60s. Today, recovering natural vegetation forms open woodland mainly dominated by *Q. alnifolia* and to a lesser extent by *Q. coccifera*. Additionally, scarce *Arbutus andrachne* and *Pinus brutia* individuals are present. Some gaps with remnants from former vineyards still exist.

In total, 207 mature individuals of *Q. alnifolia*, 66 *Q. coccifera*, and four individuals with intermediate morphological characteristics (designated hybrids) were collected from within an area of circa 5 ha (Figure 1). The sampling was exhaustive. Full-grown leaves were collected from the east side of the crown of each plant at a height of 1.50 m. Additionally, a total of 290 acorns were used for DNA analysis. These were sampled in autumn 2005 from nine maternal trees (three *Q. alnifolia*, one intermediate form and five *Q. coccifera*), which were randomly distributed within the population and partly occurred within clumping groups (Figure 1).

*Analysis of morphological data*

Discriminative leaf traits were chosen based on botanical descriptions (Meikle 1977). The protocol of leaf morphology assessment was based upon Neophytou et al. (2007), where the importance of a larger set of leaf morphological variables for hybridization studies was evaluated. Two counted variables (teeth number (TE), side vein number (UN)) and three leaf size variables (lamina length (LL), lamina width (LW), petiole length (PL) measured in cm) were assessed. All assessed variables are illustrated in Figure 2. Additionally, the following two leaf shape variables were calculated: lamina width / lamina length (LW/LL), petiole length / lamina length (PL/LL). These ratios, which form independent shape variables, have been extensively used in leaf morphometrics (Dickinson et al. 1987). Ten leaves from each plant were measured.

Multivariate analysis was employed using the statistical package R, in order to examine the simultaneous contribution of all leaf parameters in discriminating between species groups. Logarithmic transformation ($log_{10}$) was successfully employed when variables did not follow the normal distribution. Normality of the data was tested by performing Shapiro-Wilk tests for each original variable.

Principal Component Analysis (PCA) was performed in order to observe the ordination of groups of variables in principal space. The original variables were transformed to new uncorrelated variables (principal components). Discriminant Analysis was performed based on Fisher (1936), in order to verify if field assignments to taxonomic groups were in agreement with the leaf morphometric data. In particular, we aimed to classify individuals with minimum probability of misclassification. The percentage of return to original groups after the ordination in discriminant space (of their leaf data) is a measure of the true relationship of leaf variables to the taxonomic or genetic groups of plants (Pimentel 1979).

*DNA analysis*

Leaf material was first dried in vacuum and then DNA was extracted using the DNeasy 96 extraction kit (Qiagen). Subsequently, PCR was carried out for the amplification of four nuclear and seven chloroplast DNA (cpDNA) microsatellites (SSRs). Nuclear SSR locus QpZAG9 was first described in *Q. petraea* (Steinkellner et al. 1997) and loci QrZAG11, 96 and 112 in *Q. robur* (Kampfer et al. 1998). PCR programs included an initial denaturation step at 95°C lasting 8 min, 10 s of 94°C for 15 s, an annealing step lasting 15 s at 50°C (for loci QrZAG11 and QrZAG112) or 57°C (for loci QpZAG9 and QrZAG96), an elongation step at 72°C for 15 s and 23 additional cycles with reduced denaturation temperature (89°C). No final elongation was performed. The aforementioned nuclear loci were chosen after testing several SSR primer pairs from *Q. petraea* and *Q. robur* (Neophytou et al. 2010a). With regards to the cpDNA SSR loci, primers for ccmp2 were initially developed in *Nicotiana tabacum* (Weising and Gardner 1999), and for μcd4, μcd7, μdt1, μdt3 and μkk3 in *Q. petraea* and *Q. robur* (Deguilloux et al. 2003). Allele scoring was carried out by means of capillary electrophoresis using an ABI PRISM 3100 genetic analyzer (Applied Biosystems).

*Genetic diversity and differentiation based on nuclear DNA*

Measures of extant genetic diversity were calculated for both adult trees and male gametic contributions to each one of the analyzed maternal progeny arrays (effective pollen clouds). In the analysis of progenies, maternal and paternal contributions were first defined. In some cases, maternal and paternal gametic contributions were ambiguous, since mother trees and offspring shared the same heterozygous genotype. In these cases, male gametic frequency within each progeny array was estimated using a maximum likelihood method according to Gillet (1997). Subsequently, the software Rarefac (Petit et al. 1997a) was used to calculate allelic richness and gene diversity. For allelic richness rarefaction size was set to eight, corresponding to our smallest sample size. For the calculation of gene diversity, the unbiased estimator $H_k$ of Nei (1987) was used. This portion provides an estimation of the expected heterozygosity, when diploid individuals are examined. Additional measures of genetic diversity (observed heterozygosity, expected heterozygosity and null alleles at each locus) in populations of *Q. alnifolia* and *Q. coccifera* including the study stand are presented in Neophytou et al. (2010a).

Additionally, genetic differentiation among adult trees was analyzed using a multivariate approach. Factorial Correspondence Analysis (FCA; Benzécri and Bellier 1973) was used as an explorative method in order to distinguish individual grouping based on their multilocus nuclear genotypes by using the software Genetix (Belkhir et al. 2004). For the analysis, each adult genotype was converted into a three state matrix for each allele of each locus (absence of an allele is marked with 0, presence in heterozygous state is marked with 1 and presence in homozygous state is marked with 2). Independent eigenvectors of the matrix were found and individuals were placed in the factorial space.

Furthermore, male gamete heterogeneity was analyzed by means of a TwoGener analysis (Smouse et al. 2001) using the GenAlEx 6.3 software (Peakall and Smouse 2005). $\Phi_{ft}$, an analogue of Wright's $F_{ST}$, was calculated. $\Phi_{ft}$ values were computed for each pairwise comparison among effective pollen clouds of different maternal trees and their significance was tested using a non-parametric permutation procedure (10,000 permutations of male gametes between each pair of pollen clouds were applied). Analyses were carried out locus by locus and by a multilocus approach.

Finally, in order to provide a measure of the individual assignment of male gamete contributions, Bayesian analysis of genetic structures was implemented using the Structure 2.2 software (Pritchard et al. 2000). A paternity analysis was not possible with the given marker set, since paternity exclusion probability was low and assignments were ambiguous (results not shown). For the Structure analysis, allelic data of male gametic contributions to each offspring individual were used calculating posterior probabilities of membership to two assumed subpopulations or groups, without prior information of the progeny array they belonged to. In the ambiguous cases (when offspring possessed the same heterozygous genotype as the maternal tree at one locus), we assigned each allele as paternal randomly, based on the calculated frequencies in the effective pollen cloud according to the method of Gillet (1997). One-hundred-thousand burn-in periods and 100,000 Markov Chain Monte Carlo simulations were performed assuming admixture and correlated allele frequencies. Proportions of membership to the two assumed groups for each individual and population were then calculated as the means among ten runs. Two thresholds of membership proportion, $P(X|K)= 0.8$ and $P(X|K)= 0.95$, were empirically used to detect admixed haplotypes.

*Analysis of chloroplast DNA diversity and differentiation*

Chloroplast DNA haplotypes were defined as different combinations of allelic variants at the six analyzed cpDNA loci. Chlorotype nomenclature is introduced in the companion paper of Neophytou et al. (2010a) and is followed here as well. Haplotypic diversity was calculated using the software Rarefac (Petit et al. 1997a). In order to examine the degree of cpDNA sharing between the two species in our stand, introgression ratio (IG) between the two species according to Belahbib et al. (2001) was calculated. The introgression ratio was calculated as the ratio between the interspecific genetic identity and the mean of the intraspecific genetic identities. The intraspecific genetic identity equals the sum of squares of the frequencies of all haplotypic variants within each species and the interspecific one is defined as the sum of products of each haplotypic variant between species. This is described by the following formula: $IG = 2J_{12k}/(J_{1k}+J_{2k})$, where $J_{12k}$= interspecific genetic identity, $J_{1k}$, $J_{2k}$= intraspecific genetic identities. An IG score equal to zero indicates total absence of introgression between the two examined demes (in our case the two species), whereas an IG score equal to one means that all the variation is shared between them.

RESULTS

*Multivariate analysis of morphological data*

By employing Principal Component Analysis based on morphological traits, we received seven new uncorrelated synthetic variables (principal components). Among them, the first three were highly explanatory. Particularly, the first principal component explained 62.0% of the total variation, while the respective percentages were 18.2% for the second and 10.0% for the third principal component. In total, the first three principal components accounted for 90.2% of the total variation. All of them were bipolar. In terms of the first principal component the two species could be well separated. No overlapping was observed when comparing the values of the first principal component in each species. The four designated hybrids took a median position between the parental species (Figure 3). Regarding the correlation of the original variables to the synthetic ones, all leaf size variables and ratios contributed to the first principal component with loading values between 0.380 and 0.457, while the two counted variables demonstrated negative correlations (results not shown).

Using Discriminant Analysis we assigned each individual to one of the three morphological groups, corresponding to *Q. alnifolia*, *Q. coccifera* and hybrids. Percentage of return to morphological groups according to field assignment was used as a criterion for classification. All designated *Q. alnifolia* and *Q. coccifera* individuals were assigned to the respective parental morphological groups. Percentages of return to the respective morphological group were high for both species. Five designated *Q. alnifolia* individuals (2.4%) showed a percentage of return to their own morphological group lower than 0.80. For these individuals, group membership to the hybrid group varying between 0.17 and 0.40 was found. Their membership to the morphological group of *Q. coccifera* was 0.05 or less (Figure 4). Further five *Q. alnifolia* individuals presented a percentage of return to their own morphological group between 0.80 and 0.95 (result not shown). Regarding the designated *Q. coccifera* individuals, only one displayed a percentage of return to its own morphological group lower than 0.80. Two more showed a percentage between 0.80 and 0.95 (result not shown). Three designated hybrids presented a higher morphological affinity to the *Q. alnifolia* group, whilst one (individual 371) showed a higher similarity to the *Q. coccifera* group (Figure 4).

*Analysis of nuclear microsatellite data*

Large allele frequency differences between species at nuclear microsatellite loci QrZAG11 and QrZAG112 were observed. For instance, at locus QrZAG112 allele '86' was prevalent in both adults and pollen clouds of *Q. coccifera*, while allele '88' showed a very high frequency in *Q. alnifolia*. Differences of the most common alleles at each locus are presented in Figure 5. These allele frequency patterns are reflected into the diversity parameters calculated for the respective loci. Thus, in all interspecific comparisons diversity ($H_k$) was significantly different (Table 1). On the other hand, we found no significant differences between adults and pollen clouds within the same species for any of the studied loci, which supports that interspecific pollinations might be very limited. Interestingly, in the case of the hybrid mother tree, male gametes showed a high affinity to *Q. alnifolia*. Diversity was not significantly different between the pooled pollen clouds of *Q. alnifolia* progenies and the pollen cloud of the hybrid progeny for any of the studied loci. In contrast, it was significantly different in comparison to the pollen clouds of the *Q. coccifera* progenies for all loci (Table 1).

By employing Factorial Correspondence Analysis on genotypic data, we received four new uncorrelated synthetic variables (factors). 61.85% of the total explained variation was found in the first two factors. Species could be well distinguished in factorial space and overlapping was very limited (Figure 6). *Q. alnifolia* formed a markedly more tightly bulked group, which reflects its lower genetic variation in comparison to *Q. coccifera*. The four designated hybrids were positioned between the parental species (Figure 6). A high relative contribution to the first factor was observed among all diagnostic alleles. In particular, allele '88' of locus QrZAG112 showed the highest relative contribution followed by alleles '251' of locus QrZAG11 and '86' of locus QrZAG112 (results not shown).

*cpDNA diversity and differentiation*

Analysis of cpDNA microsatellites confirmed low levels of introgression between the two species. In total, six cpDNA haplotypes (chlorotypes) were observed in the mixed stand and were named according to Neophytou et al. (2010a). *Quercus alnifolia* was markedly less diverse than *Q. coccifera* ($H_k$= 0.136 versus $H_k$= 0.651 respectively; Table 2). Among the three observed chlorotypes of *Q. alnifolia,* the most common, chlorotype 7, occurred in 93% of the individuals and chlorotype 6 had a frequency of 7% in this species (Figure 7). Chloroplast DNA sharing was very low. Chlorotype 7 was found in 13% of *Q. coccifera* trees and chlorotype 6 was found in just one *Q. coccifera* tree (1.5%). The remaining chlorotypes were species-specific. In particular, chlorotypes 5, 11 and 12 were confined to *Q. coccifera* and chlorotype 13 was a rare variant observed in only one *Q. alnifolia* individual (Figure 7). Finally, the introgression ratio (IG) was limited to 0.026 reflecting the low degree of cytoplasmic sharing between the two species.

*Comparison of genetic and morphological assignments*

In order to compare the morphological and genetic admixture among adult individuals, results of a Structure assignment analysis of genotypes carried out in the companion paper of Neophytou et al. (2010a) were additionally used. Individuals were chosen, which possessed membership proportion lower than 0.80 to their own species group either based on the Discriminant Analysis (morphological assignment) of on Structure Analysis (genetic assignment). A comparison was made between genetic and morphological assignments of these individuals (Figure 4). The four designated hybrids were also included to this comparison. Finally, chlorotypes were given for all aforementioned individuals.

None of the designated parental species individuals displayed both morphological and genetic admixture. In particular, morphologically admixed *Q. alnifolia* individuals displayed P(X|K) higher than 0.97, whilst the single morphologically admixed *Q. coccifera* individual showed P(X|K)= 0.91 to its own species cluster. In contrast, all genetically admixed parental species individuals presented a high percentage of return (at least 0.99) to their own species, based on the Discriminant Analysis. Two designated hybrids showed admixed morphology and genotypes (individuals 293 and 371), and two of them showed morphological affinity to *Q. alnifolia*, but genetic affinity to *Q. coccifera* (individuals 348 and 352). None of the individuals with admixed morphological or genotypic assignment possessed a heterospecific chlorotype (Figure 4). Designated hybrids possessed chlorotype 5 (typical in *Q. coccifera*) in three cases and chlorotype 7 (typical in *Q. alnifolia*) in one (Figure 4). Use of 0.95 as a threshold of membership proportion did not change the outcome of this comparison, since morphological and genetic admixture did not coincide (results not shown).

*Genetic differentiation among effective pollen clouds*

High differentiation among male gametes of interspecific progeny arrays was observed. Multilocus $\Phi_{ft}$ values varied between 0.306 and 0.453 and were all highly significant (P<0.001; Table 3). The corresponding values for the comparisons within species were much lower: 0.053-0.094 among pollen clouds of *Q. alnifolia* mother trees and 0.000-0.137 among pollen clouds of *Q. coccifera* mother trees. Most of the comparisons were significant at the P<0.05 or P<0.01 level. The higher differentiation values in *Q. coccifera* were due to the progeny array of mother tree C349 which differed significantly from all other intraspecific mothers in terms of male gametic diversity. By removing this progeny array, $\Phi_{ft}$s in *Q. coccifera* reached a maximum of 0.047 being of the same magnitude as in *Q. alnifolia* (Table 3). A further aspect of genetic differentiation among pollen clouds is the genetic structure within species, which seems to be linked with mother tree location. For instance, male gametes that pollinated neighbouring mother trees C282 and C283 (both *Q. coccifera*; Figure 1) do not differ from each other significantly. In general, results for both species reveal genetic structures due to heterogeneous male contributions among mother trees with $\Phi_{ft}$ values among groups being significant (results not shown). The pollen cloud of the hybrid mother tree showed a much higher affinity to *Q. alnifolia*. Two out of three comparisons to *Q. alnifolia* were non-significant, whilst all comparisons to *Q. coccifera* were highly significant (Table 3).

Genetic assignment analysis using the software Structure provided further insights into differentiation patterns within and between effective pollen clouds. Effective pollen clouds of *Q. alnifolia* mother trees were highly homogeneous with membership proportion to the respective species cluster (P(X|K)) varying between 0.942 and 0.970 (Table 4). Individual membership proportions were high. Male gametes of 70 progenies out of 71 were sorted to the *Q. alnifolia* cluster with P(X|K)>0.8. Fifty-seven of them showed P(X|K)>0.95. In one case, the paternal tree displayed a higher degree of admixture (0.5<P(X|K)<0.8), but no male gamete was assigned to the *Q. coccifera* cluster. Among *Q. coccifera* progeny arrays results were more heterogeneous. Male gametes of 165 progenies out of 176 were sorted to the *Q. coccifera* cluster with P(X|K)>0.8. Among them, 53 possessed membership proportions between 0.8 and 0.95. As revealed by P(X|K), for *Q. coccifera* maternal tree C349 among 14 male gametes, four were assigned to *Q. alnifolia* (two with 0.95>P(X|K)>0.8 and two with P(X|K)>0.95; Table 4). Another similar case was observed in *Q. coccifera* maternal tree C367 (one male gamete was assigned to the *Q. alnifolia* cluster with 0.95>P(X|K)>0.8). Regarding the hybrid progeny array, group membership proportion of male gametes to the derived *Q. alnifolia* cluster was 0.884 indicating a prevalence of *Q. alnifolia* within the effective pollen cloud. Out of 43 analyzed male gametes, 38 were sorted to the *Q. alnifolia* cluster with P(X|K)>0.8 and two were sorted to the *Q. coccifera* cluster with 0.95>P(X|K)>0.8 (Table 4).

DISCUSSION

The present study aimed to focus on the analysis of intra- and interspecific differentiation and gene flow in a single sympatric population of *Quercus alnifolia* and *Quercus coccifera*, by using various statistical approaches. Results widely agree that hybridization between the two species is very limited. In a total of 277 adult individuals, 273 were recognized as 'pure' *Q. alnifolia* or *Q. coccifera* in the field (207 *Q. alnifolia* and 66 *Q. coccifera*). Some of them were assigned as intermediate by the morphological analysis and some of them possessed an admixed genotype as revealed by a previous Structure analysis (Neophytou et al. 2010a). However, none of them could be characterized as an intermediate by both analyses. The two species were also well separated in terms of chloroplast DNA haplotypes. The notion of very limited hybridization is further supported by analysis of male gametic contributions to the progenies of selected maternal trees (effective pollen clouds) which reveals that levels of successful interspecific crossings are very low.

The multivariate analysis of morphological traits has been proved once again to be a powerful tool for distinguishing between *Q. alnifolia* and *Q. coccifera*. Results from the mixed stand of the present study agree to a large extent with past findings including pure populations and transects with sympatric stands from Troodos Mountains (Neophytou et al. 2007). Although we used a reduced number of original variables, based on leaf morphometric traits, we received an even higher percentage of explained variation by the first three principal components, which exceeded 90%. This confirms the validity of the chosen morphological traits for discriminating between the two species. Even using only the first principal component, we received a clear-cut separation between *Q. alnifolia* and *Q. coccifera*. Both in the present study and in Neophytou et al. (2007) the two species are well distinguished and overlapping is limited, supporting that introgression is low, at least arising from morphological traits. A further result which agrees with the previous findings is that hybrids show a closer morphological affinity to *Q. alnifolia*. None of the four intermediate individuals in our study occurs within the cluster of *Q. coccifera*. On the scatter plot of the PCA, three of them appear within the *Q. alnifolia* cluster and one possesses a median position between the two species' clusters.

Results from the analyses of nuclear microsatellites provided further evidence that interspecific introgression is low. Additionally, multilocus estimates of genetic variation at nuclear SSRs, as well as cpDNA haplotype diversity were lower in *Q. alnifolia*. This is probably due to the different evolutionary history of the two species. Toumi and Lumaret (2001) in an alloenzyme study including *Q. alnifolia*, *Q. coccifera* and other related sclerophyllous oak species found that *Q. alnifolia* possessed the lowest diversity values among all species studied and attributed this to a founder effect. On one hand, *Q. coccifera* has a wide distribution across the Mediterranean. This species and its ancestral species, *Q. mediterranea*, had a strong presence in this area at least during the Oligocene (Palamarev 1989). On the other hand, *Q. alnifolia* is endemic and confined to a limited area on the igneous rocks of Troodos Mountains on Cyprus. Fossil evidence of its ancestral species, *Q. pseudoalnus*, is limited (Palamarev 1989). Thus, it might have been to a greater extent subjected to genetic drift effects during its evolutionary history.

Genetic drift may also account for the high allele frequency differences between the two species at loci QrZAG11 and 112. The aspect of high nuclear genetic differentiation and low admixture between the two species has already been discussed in Neophytou et al. (2010a). In the present study, we extended the analyses by studying genetic diversity and differentiation among progeny arrays from known maternal trees of both species, as well as of a designated hybrid. Genetic differentiation between effective pollen clouds is very high and significant between all interspecific pairs of mother trees. At the within species level, differentiation among pollen clouds was lower, but in several cases significant, which may indicate assortative mating. Heterogeneity among pollen pools received by different maternal trees of the same species has been observed in other oak species as well and has been attributed to within population differences in flower maturation timing (e.g. Streiff et al. 1999 for *Q. petraea* and *Q. robur*).

More detailed information about the frequency of interspecific crossings was revealed by the Bayesian clustering analysis of male gametes. In particular, no evidence for interspecific pollinations was found for any individual among *Q. alnifolia* progeny arrays. Among progeny arrays of *Q. coccifera* mother trees, there was evidence for only five interspecific matings out of a total of 176 analyzed offspring (2.8%). Four of them occurred in a specific maternal tree (C349; 28.6 % of the total male gametic contributions to the progeny of this tree). The frequency of interspecific pollinations between *Q. alnifolia* and *Q. coccifera* appears to be lower in comparison to the well studied European white oaks. For instance, Salvini et al. (2009) found that 26% of *Q. pubescens* offspring (acorns) were fathered by *Q. petraea* in a mixed stand of the two species in Italy. In the case of *Q. robur* and *Q. petraea* this percentage varied between 17 and 48% among different progenies of the first species, whereas the second species acted almost exclusively as pollen donor (Bacilieri et al. 1996).

A directionality of interspecific pollen flow was observed in our data, with *Q. alnifolia* acting as pollen donor. The hybrid mother tree was mainly pollinated by *Q. alnifolia*, whilst we could not detect any *Q. coccifera* paternal contribution among progenies of *Q. alnifolia* mother trees. An important factor playing a role in pollination patterns is the relative abundance and spatial distribution of the two species (Lepais et al. 2009, Varela et al. 2008). In our study stand, *Q. alnifolia* is predominant (Figure 1), which could be the reason of the genetic affinity of its effective pollen cloud to the '*Q. alnifolia*' cluster. Additionally, neighbouring trees C282 and C283 possessed effective pollen clouds with minimal differentiation. Finally, for all interspecific pollinations of *Q. coccifera* tree C349, which demonstrated the highest degree of interbreeding among all mother trees, the pollen donor might have been a nearby *Q. alnifolia* (A351, at 2 m distance), as its genotype matches the male gametic contribution of all interbred offspring. However, we note here that for each one of the interbred offspring of this tree, up to three adult *Q. alnifolia* trees of the study stand possessed a matching genotype and could have been potential male parents. Thus we were not able to assign the father unambiguously. The analysis of additional SSRs is necessary to obtain higher exclusion probabilities in paternity analyses.

The rarity of interspecific crossings indicates that effective reproductive barriers might act between the two species preventing successful interspecific mating. Flowering periods of the two species are widely overlapping with *Q. alnifolia* flowering between the end of April and the beginning of June (Knopf 2006), whereas dates for *Q. coccifera* span from April to May (Chatziphilippidis 2006). This gives the opportunity of interspecific pollinations. Observations from year 2009 in our study stand confirmed overlapping in flowering. However, in *Q. coccifera* flower maturation was slightly more advanced. The hybrid mother tree showed a delay in flowering. Given that the two species are protandrous, this asynchrony could have resulted in a higher degree of fertilization of *Q. coccifera* and hybrid female flowers with *Q. alnifolia* pollen (it is more likely that female flowers of *Q. coccifera* are receptive during maturation of male flowers of *Q. alnifolia* than the opposite). In the case of the interfertile *Q. ilex* and *Q. suber*, flowering timing has been proposed to affect hybridization directionality with the latter species flowering later and acting almost exclusively as pollen donor (Varela et al. 2008). Apart from flowering phenology, post-pollination mechanisms may also prevent interbreeding. Pollen-pistil interactions may set a prezygotic barrier by inhibiting of pollen tube growth and embryo formation. In addition, even if fertilization is successful, acorn maturation may be incomplete and immature seeds from interspecific crosses may be aborted (Boavida et al. 2001). Given that *Q. alnifolia* and *Q. coccifera* show differences in their reproductive cycles (annual acorn maturation in *Q. alnifolia* and predominantly biennial in *Q. coccifera*; Knopf 2006, Bianco and Schirone 1985), such physiological incompatibilities possibly act preventing hybridization between the two species.

Results from cpDNA microsatellites further confirm very low levels of introgressive hybridization between *Q. alnifolia* and *Q. coccifera* in the study stand. A limited sharing was observed mainly concerning haplotype 7, dominant in *Q. alnifolia* of this stand, which was found in eight *Q. coccifera* trees (12%). Nonetheless, there is no evidence that this was due to recent hybridization since genotypes of these individuals were not introgressed (membership proportions to *Q. coccifera* cluster 0.972-0.990 based on Structure analysis in Neophytou et al. (2010a); results not shown). Moreover, since no sequence information from the analyzed cpDNA SSR loci was available, size homoplasy cannot be ruled out.

This lack of cpDNA introgression at a local scale revealed by the present study may indicate a relatively recent contact of the two species at this certain study site. On the other hand, large scale cpDNA introgression revealed by multipopulation data (Neophytou et al. 2010a) might be due to historical hybridization events. The spatial distribution of cpDNA lineages in other sclerophyllous oaks of the Mediterranean reflects ancient migration patterns, since these species persisted in the area throughout the Pleistocene (e.g. Jiménez et al. 2004; López de Heredia et al. 2007). Large scale chloroplast DNA sharing between *Q. alnifolia* and *Q. coccifera* is probably the imprint of rare hybridization events that sporadically happened during their long existence on Cyprus.

ACKNOWLEDGMENTS

We would like to thank Andreas Christou (Department of Forests, Republic of Cyprus personnel, for their assistance during plant collections). We are grateful to Mr. Petros Anastasiou for his valuable help during plant collections. This research was conducted in partial fulfilment for the doctorate degree of the first author at Albert-Ludwigs University of Freiburg. During that time the first author was supported by doctorate scholarships from DAAD and the State of Baden-Württemberg. Partial financial assistance to Filippos Aravanopoulos in the form of two cooperative grants of the Ministry of Natural Resources of Cyprus and the Aristotle University of Thessaloniki is gratefully acknowledged.

TABLES

Table 1 – Nuclear genetic diversity measures of adult trees and male gametes of the analyzed progenies (effective pollen clouds) in each species. N= number of analyzed chromosomes (2n for the adults and n for the pollen clouds), $n_{al}$= number of alleles, $H_k$= Nei's unbiased gene diversity, $SE_{Hk}$= standard error of Nei's unbiased gene diversity and $R_8$= allelic richness (rarefaction size= 8 haploid individuals).

| Locus | | Adults | | | Pollen clouds | | |
|---|---|---|---|---|---|---|---|
| | | Q. alnifolia | Q. coccifera | Hybr. | Q. alnifolia | Q. coccifera | Hybr. |
| QpZAG9 | N | 414 | 132 | 8 | 57 | 172 | 43 |
| | $n_{al}$ | 11 | 12 | 4 | 7 | 14 | 9 |
| | $H_k$ | 0.722 | 0.858 | 0.821 | 0.736 | 0.850 | 0.798 |
| | $SE_{Hk}$ | 0.015 | 0.014 | 0.101 | 0.035 | 0.012 | 0.037 |
| | $R_8$ | 3.781 | 5.136 | 4.000 | 3.754 | 5.038 | 4.422 |
| QrZAG11 | N | 412 | 130 | 8 | 55 | 170 | 42 |
| | $n_{al}$ | 6 | 13 | 4 | 2 | 13 | 2 |
| | $H_k$ | 0.157 | 0.822 | 0.643 | 0.105 | 0.805 | 0.048 |
| | $SE_{Hk}$ | 0.024 | 0.020 | 0.184 | 0.055 | 0.019 | 0.045 |
| | $R_8$ | 1.613 | 4.788 | 4.000 | 1.382 | 4.620 | 1.190 |
| QrZAG96 | N | 414 | 132 | 8 | 69 | 171 | 33 |
| | $n_{al}$ | 5 | 5 | 2 | 3 | 6 | 3 |
| | $H_k$ | 0.520 | 0.412 | 0.250 | 0.571 | 0.339 | 0.595 |
| | $SE_{Hk}$ | 0.023 | 0.048 | 0.180 | 0.043 | 0.045 | 0.043 |
| | $R_8$ | 2.617 | 2.415 | 2.000 | 2.640 | 2.277 | 2.571 |
| QrZAG112 | N | 414 | 132 | 8 | 68 | 175 | 43 |
| | $n_{al}$ | 7 | 4 | 4 | 3 | 3 | 3 |
| | $H_k$ | 0.424 | 0.117 | 0.750 | 0.403 | 0.109 | 0.451 |
| | $SE_{Hk}$ | 0.026 | 0.038 | 0.139 | 0.054 | 0.031 | 0.080 |
| | $R_8$ | 2.412 | 1.462 | 4.000 | 1.031 | 1.396 | 2.454 |
| Multilocus | $n_{al}$ | 7.250 | 8.500 | 3.500 | 3.750 | 9.000 | 4.250 |
| | $H_k$ | 0.456 | 0.552 | 0.616 | 0.454 | 0.526 | 0.473 |
| | $SE_{Hk}$ | 0.235 | 0.354 | 0.255 | 0.302 | 0.361 | 0.317 |
| | $R_8$ | 2.606 | 3.450 | 3.500 | 2.202 | 3.333 | 2.659 |

Table 2 – Haplotypic diversity and differentiation between *Q. alnifolia* and *Q. coccifera*. measures of adult trees and male gametes of the analyzed progenies (effective pollen clouds) in each species. N= number of analyzed individuals, $n_{hap}$= number of haplotypes, $H_k$= Nei's unbiased gene diversity, $SE_{Hk}$= standard error of Nei's unbiased gene diversity.

|  | *Q. alnifolia* | *Q. coccifera* | Hybr. |
|---|---|---|---|
| N | 207 | 66 | 4 |
| $n_{hap}$ | 3 | 5 | 2 |
| $H_k$ | 0.136 | 0.651 | 0.521 |
| $SE_{Hk}$ | 0.031 | 0.035 | 0.265 |

Table 3 – Multilocus pairwise Φft values among male gametes of each analyzed progeny array. Mother trees: A= *Quercus alnifolia*, C= *Quercus coccifera*, H= intermediates. Locations of the trees are shown in Figure 1.

|  | A216 | A290 | A441 | C282 | C283 | C349 | C361 | C367 |
|---|---|---|---|---|---|---|---|---|
| A290 | 0.053* | | | | | | | |
| A441 | 0.053 | 0.094* | | | | | | |
| C282 | 0.348*** | 0.453*** | 0.385*** | | | | | |
| C283 | 0.343*** | 0.446*** | 0.372*** | 0.000 | | | | |
| C349 | 0.268*** | 0.378*** | 0.306*** | 0.137*** | 0.128*** | | | |
| C361 | 0.342*** | 0.409*** | 0.356*** | 0.047** | 0.033* | 0.052** | | |
| C367 | 0.360*** | 0.440*** | 0.386*** | 0.036** | 0.019 | 0.111*** | 0.019* | |
| H371 | 0.023 | 0.058*** | 0.043 | 0.307*** | 0.297*** | 0.204*** | 0.278*** | 0.312*** |

Table 4 – Proportions of membership of male gametes to species group
Results based on the Structure genetic assignment analysis are presented. Mother trees, species and number of sampled offspring (acorns) per tree are given in the first three columns. Proportions of membership of each effective pollen cloud to each one of the two inferred clusters are given in the fourth column (P(X|K)). Two threshold values of P(X|K) were set; 0.8 and 0.95. Number of individuals is given for each class of membership proportion; e.g. N (0.95<P<0.8) for individuals showing 0.95<P(X|K)<0.8 etc.

|  |  |  | *Quercus alnifolia* cluster | | | | *Quercus coccifera* cluster | | | |
|---|---|---|---|---|---|---|---|---|---|---|
|  |  |  | Group member-ship | Number of individuals per class of membership proportion | | | | | | Group member-ship |
| Moth. tree | Sp. | N | P(X\|K) | N (P>0.95) | N (0.95<P<0.8) | N (0.5<P<0.8) | N (0.5<P<0.8) | N (0.8<P<0.95) | N (P>0.95) | P(X\|K) |
| A216 |  | 15 | 0.970 | 14 | 1 | 0 | 0 | 0 | 0 | 0.031 |
| A290 | *Q.* | 48 | 0.956 | 37 | 10 | 1 | 0 | 0 | 0 | 0.044 |
| A441 | *alnifolia* | 8 | 0.942 | 6 | 2 | 0 | 0 | 0 | 0 | 0.058 |
| Total |  | 71 | 0.957 | 57 | 13 | 1 | 0 | 0 | 0 | 0.043 |
| H371 | Hybrid | 43 | 0.884 | 30 | 8 | 1 | 2 | 2 | 0 | 0.116 |
| C282 |  | 34 | 0.055 | 0 | 0 | 0 | 0 | 6 | 28 | 0.945 |
| C283 |  | 28 | 0.106 | 0 | 0 | 0 | 3 | 6 | 19 | 0.894 |
| C349 | *Q.* | 14 | 0.337 | 2 | 2 | 0 | 2 | 5 | 3 | 0.663 |
| C361 | *coccifera* | 50 | 0.075 | 0 | 0 | 0 | 0 | 25 | 25 | 0.925 |
| C367 |  | 50 | 0.070 | 0 | 1 | 0 | 1 | 11 | 37 | 0.930 |
| Total |  | 176 | 0.095 | 2 | 3 | 0 | 6 | 53 | 112 | 0.905 |

FIGURES

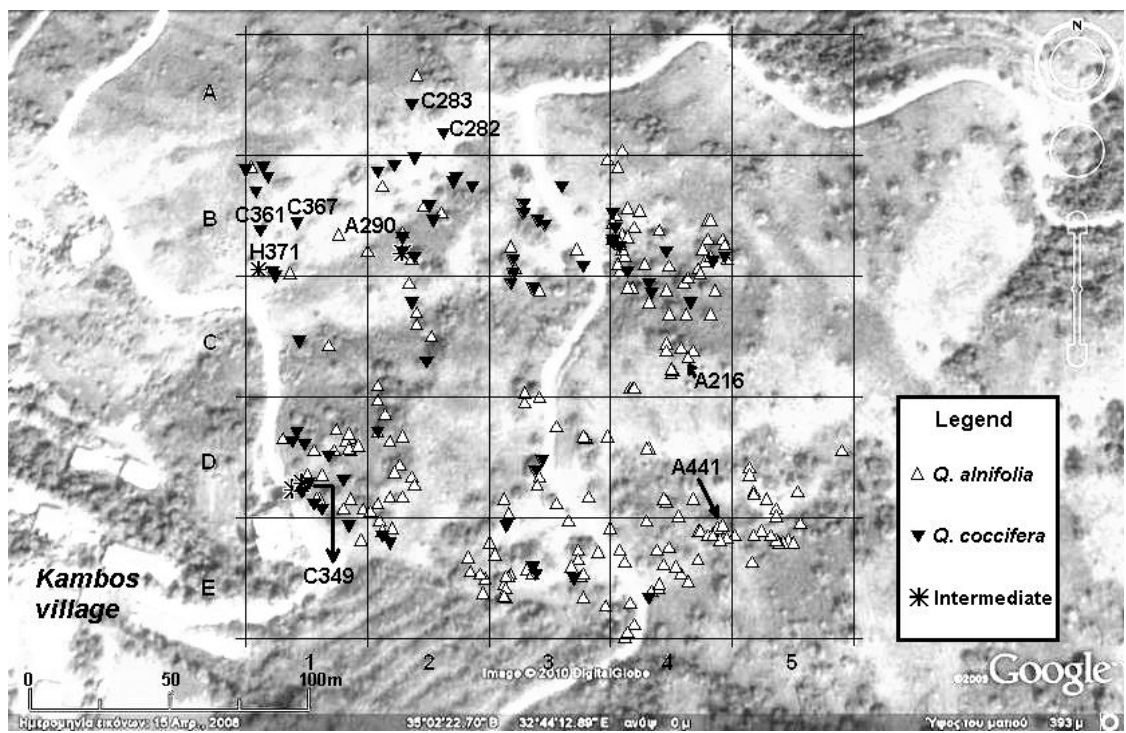

Figure 1 – Satellite image of the mixed stand (retrieved by Google Earth). All adult trees belonging to the two parental species are presented. The mother trees corresponding to the analyzed progeny arrays are also located (Species is given as prefix of tree number A= *Q. alnifolia*, C= *Q. coccifera*, H= designated hybrid).

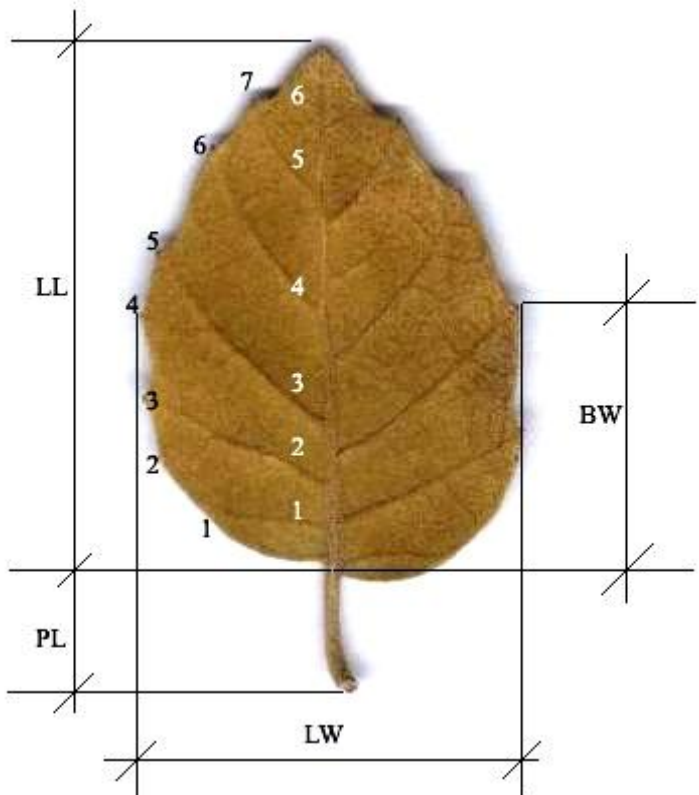

Figure 2 – Illustration of the assessed leaf morphological parameters. Black numbers denote teeth number (TE) and white numbers denote side vein number (UN). LL= lamina length, LW= lamina width, PL= petiole length (PL).

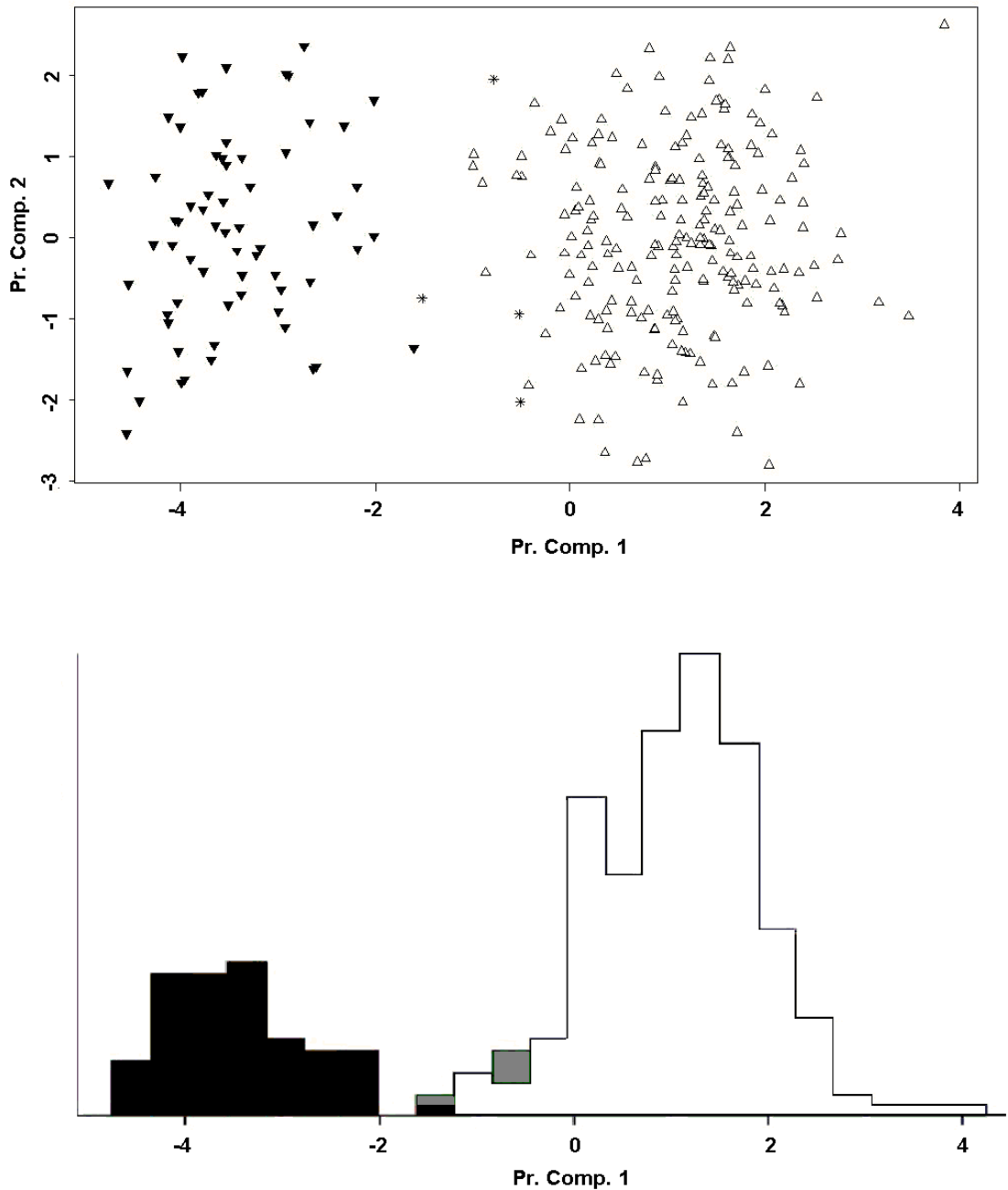

Figure 3 – Principal Component Analysis (PCA) of leaf morphological traits. A scatter plot of the first two principal components and a frequency distribution bar plot based on the first principal component are included. *Quercus alnifolia* is denoted with non-filled triangles on the scatter plot and non-filled columns on the bar plot. *Quercus coccifera* is denoted with black-filled inverse triangles on the scatter plot and black-filled columns on the bar plot. Putative hybrids are denoted with asterisks on the scatter plot and grey-filled columns on the bar plot.

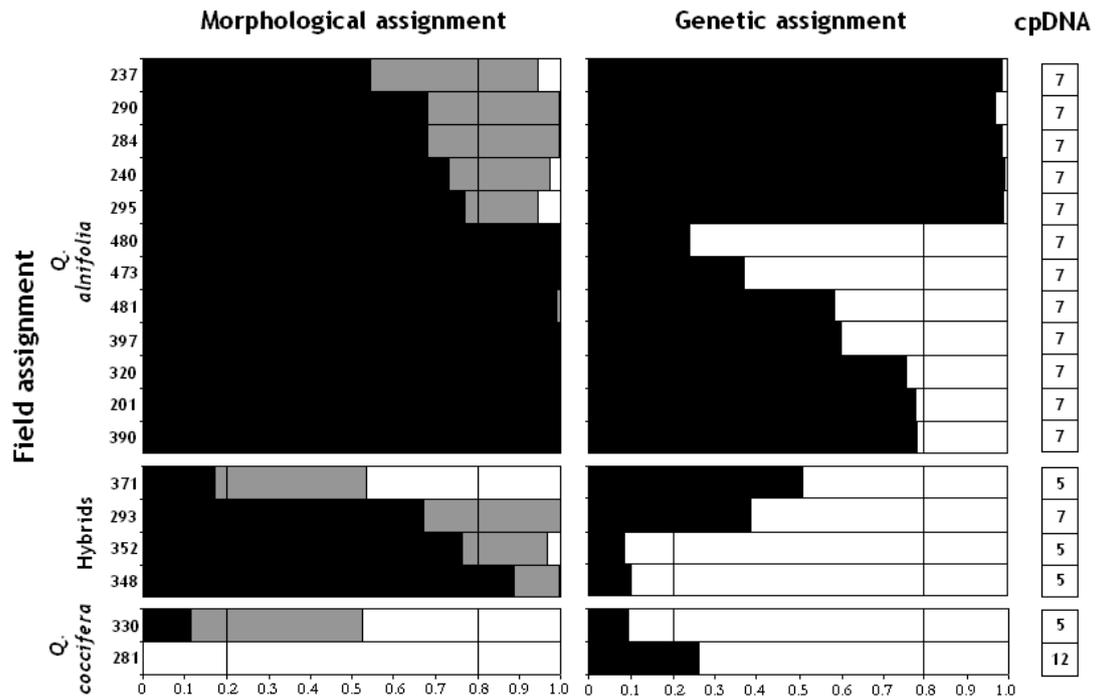

Figure 4 – Results of the discriminant analysis of leaf morphometric characters and the Structure analysis of nuclear microsatellites (Neophytou et al. 2010a) are presented with bar plots for individuals with membership proportions to their own cluster lower than 0.8 (revealed either by the morphological or by the genetic analysis). Membership proportion to each of the derived clusters is shown. Black colour is used for the derived cluster of *Q. alnifolia*, grey for hybrids (only in the Discriminant Analysis) and white for *Q. alnifolia*. Chlorotype of each individual is given in a separate column.

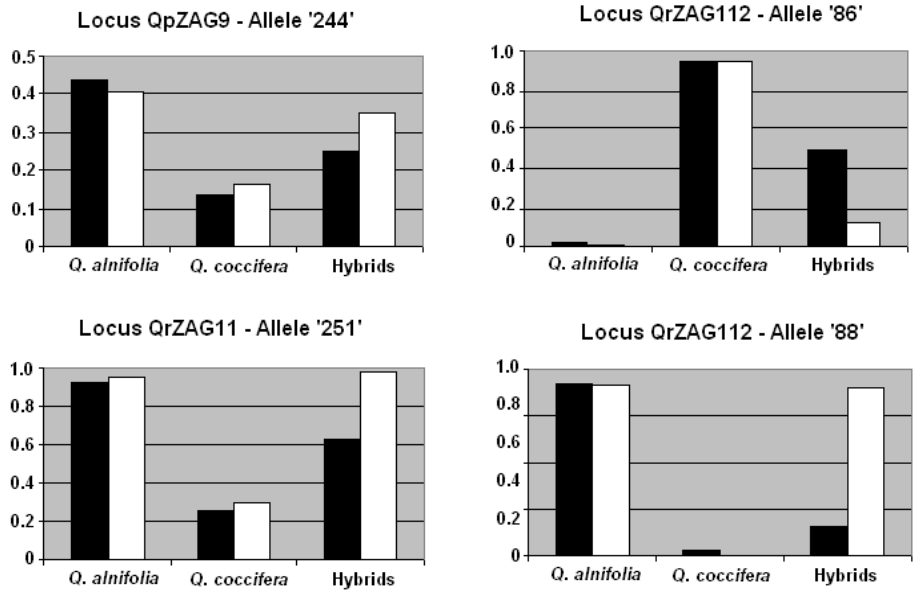

Figure 5 – Distribution of diagnostic alleles among adult trees (black-filled) and progenies (white-filled) in *Q. alnifolia*, *Q. coccifera* and putative hybrids.

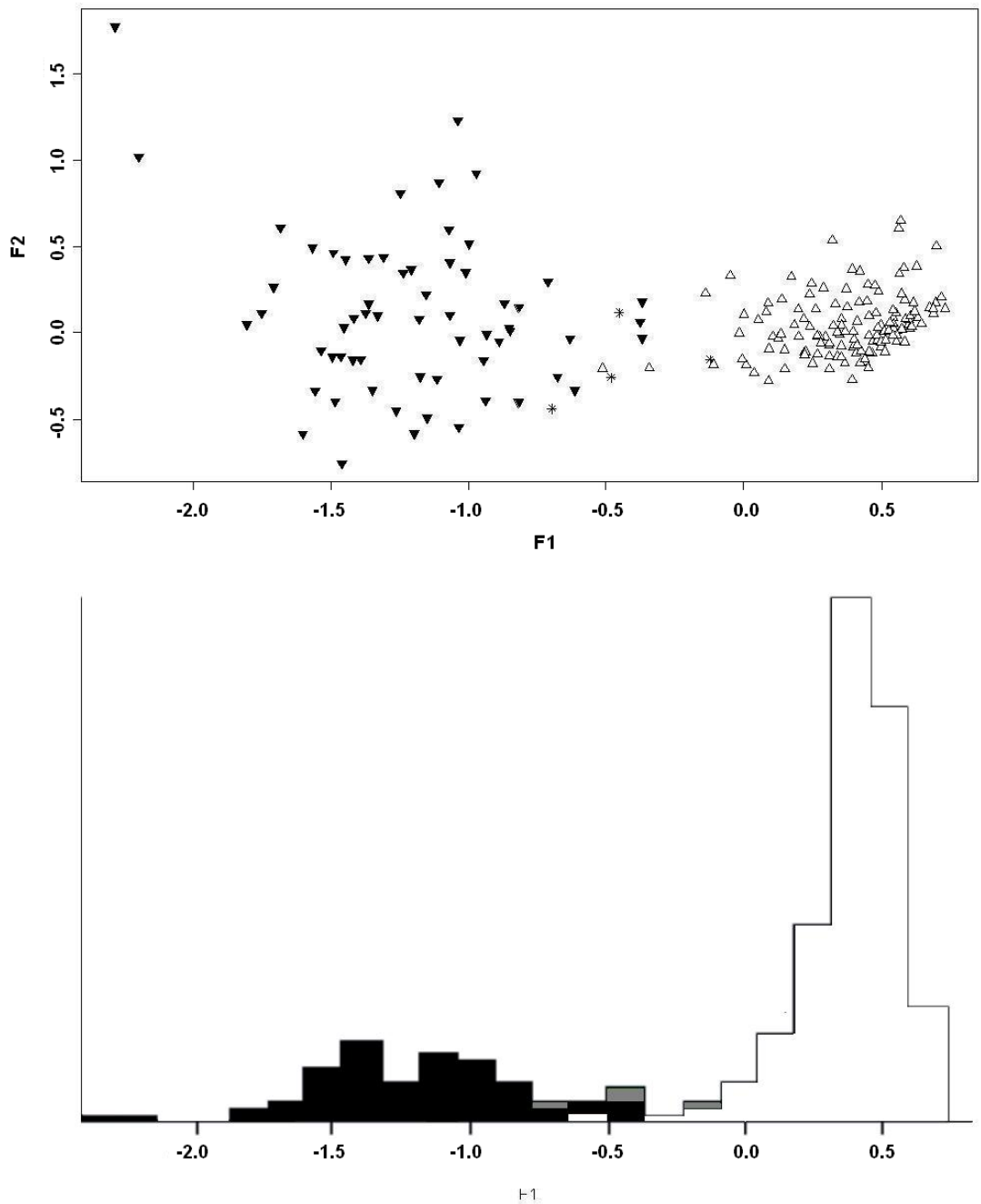

Figure 6 – Factorial Correspondence Analysis (PCA) of nuclear microsatellite data. A scatter plot of the first two factors and a frequency distribution bar plot based on the first factor are included. *Quercus alnifolia* is denoted with non-filled triangles on the scatter plot and non-filled columns on the bar plot. *Quercus coccifera* is denoted with black-filled inverse triangles on the scatter plot and black-filled columns on the bar plot. Putative hybrids are denoted with asterisks on the scatter plot and grey-filled columns on the bar plot.

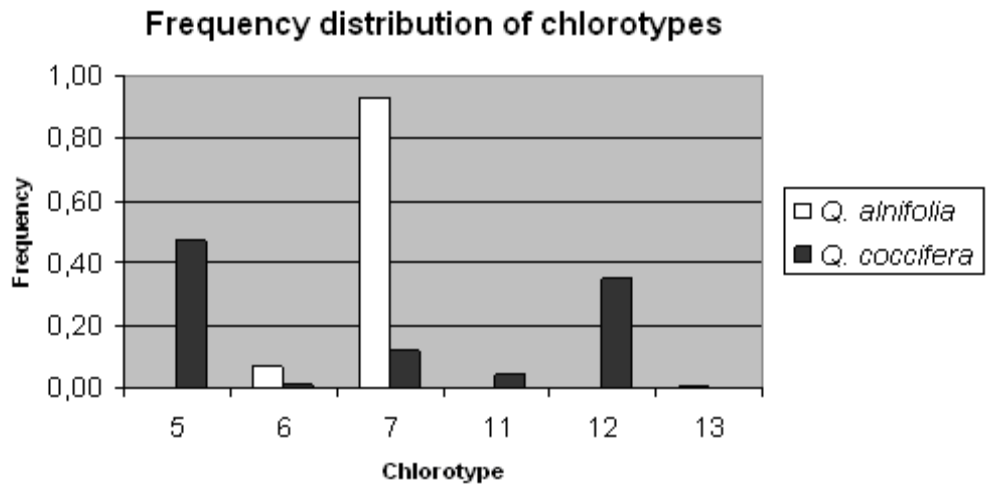

Figure 7 – Frequency distribution of chloroplast DNA haplotypes. Chlorotype designation is based upon Neophytou et al. (2010a).